\documentclass[10pt,conference]{IEEEtran} 
\IEEEoverridecommandlockouts

\usepackage{amsmath,amssymb,amsfonts}
\usepackage{algorithmic}
\usepackage{graphicx}
\usepackage{textcomp}
\usepackage{xcolor}
\usepackage[noadjust]{cite}
\usepackage{braket}
\usepackage{array}
\usepackage{mathtools}
\usepackage{qcircuit}

\begin{document}
\title{Modeling Time-Dependent Systems using Dynamic Quantum Bayesian Networks}

\author{\IEEEauthorblockN{Sima E. Borujeni}
\IEEEauthorblockA{Department of Industrial, Systems,\\ and Manufacturing Engineering\\
Wichita State University\\
Wichita, Kansas 67260\\
Email: sxborujeni@shockers.wichita.edu}
\and
\IEEEauthorblockN{Saideep Nannapaneni}
\IEEEauthorblockA{Department of Industrial, Systems,\\ and Manufacturing Engineering\\
Wichita State University\\
Wichita, Kansas 67260\\
Email: saideep.nannapaneni@wichita.edu}}

\maketitle
\begin{abstract}

Advances in data collection using inexpensive sensors have enabled monitoring the performance of dynamic systems, and to implement appropriate control actions to improve their performance. Moreover, engineering systems often operate under uncertain conditions; therefore, the real-time decision-making framework should not only consider real-time sensor data processing but also several uncertainty sources that may impact the performance of dynamic systems. In this paper, we investigate the modeling of such time-dependent system behavior using a dynamic quantum Bayesian network (DQBN), which is the quantum version of a classical dynamic Bayesian network (DBN). The DBN framework has been extensively used in various domains for its ability to model stochastic relationships between random variables across time. The use of the quantum amplitude amplification algorithm provides quadratic speedup for inference and prediction in Bayesian networks. In this paper, we combine the modeling capabilities of DBN with the computational advantage of quantum amplitude amplification for efficient modeling and control of time-dependent systems. We implement the proposed DQBN framework on IBM Q hardware, and compare its performance with classical DBN implementation and the IBM Qiskit simulator.

\end{abstract}
\begin{IEEEkeywords}
Dynamic, Control, Bayesian, Quantum, Qiskit, IBM, Experimental, Circuit.
\end{IEEEkeywords}
\IEEEpeerreviewmaketitle

%%% -----------------------------------------------------------------
\section{Introduction} \label{Intro}

Time-dependent systems are those systems whose performance characteristics change over time. A typical example of such a system is an engineering system (e.g. load-bearing structural system such as civil infrastructure \cite{luque2019risk} and aircraft system \cite{li2017dynamic}) whose performance varies due to underlying mechanical degradation over time. In addition to such engineering systems, time-dependent system behavior is also observed in financial systems (stock markets) \cite{liu2019dynamical}, healthcare (patient health management) \cite{van2008dynamic}, robotics \cite{premebida2017dynamic}, speech recognition \cite{nefian2002dynamic}, and bioinformatics \cite{zou2005new}. Due to the dynamic nature of the system's performance, it becomes essential to monitor its performance in real time to ensure that the system meets the operational requirements. Early detection of any potential faults is desirable as such faults may lead to system failures leading to safety and economic consequences \cite{jiang2018data}. 

Such real-time system performance monitoring falls into the \textit{digital twin} paradigm where a computer model is used as a twin to the physical system, and this computer model can be used for real-time decision-making to minimize the effects of any failures and this improve a system's operational health \cite{vanderhorn2021digital, kapteyn2020physics}.

Engineering systems often operate under uncertain conditions; such uncertainty sources need to be identified and included in a comprehensive modeling framework \cite{nannapaneni2016uncertainty}. In the context of structural systems, these uncertainty sources can include the variation in the load on the system, uncertainty in the material properties, and uncertainty in the amount of degradation in the system \cite{li2017dynamic}. One of the widely used frameworks for modeling time-dependent systems is a Dynamic Bayesian Network (DBN), which is an extension of a Bayesian network (BN) for modeling a dynamic system. A BN has the inherent capability to represent uncertain variables, and thus, a DBN is used to represent stochastic relationships between variables over time.

Performance monitoring and fault detection is modeled as an inference problem (i.e. inferring the state of a system from observation data), and quantum algorithms have been shown to have superior computational performance over classical implementation. One of the earliest algorithms is the quantum rejection sampling algorithm, which uses the quantum amplitude amplification algorithm through the Grover's search operator \cite{grover1996fast} to achieve a quadratic speedup in the inference analysis \cite{ozols2013quantum, low2014quantum}. Our prior work developed a generalized discrete Quantum Bayesian network (QBN) framework on a gate-based quantum computing platform \cite{borujeni2021quantum}, and we performed an experimental evaluation of the accuracy of QBN analysis on various IBM hardware and compared their performance to the IBM Qiskit simulator and classical implementation \cite{borujeni2020experimental}.

In this paper, we develop a Dynamic Quantum Bayesian Network (DQBN) framework that integrates the computational benefits of inference analysis in QBN with the capabilities of DBN in modeling time-dependent systems for real-time system performance monitoring. We discuss the DQBN circuit representation, inference analysis, and study the solution accuracy when implemented on IBM hardware, ibmq\_16\_melbourne and the IBM Qiskit simulator against classical implementation \cite{ibmq}. 

\textbf{Paper Organization:} The rest of the paper is organized as follows. Section \ref{bg} provides a brief background to Dynamic Bayesian networks and Quantum Bayesian networks. Section \ref{sec:method} discusses the proposed Dynamic Quantum Bayesian network framework by integrating the principles of Dynamic and Quantum Bayesian networks. Section \ref{sec:case} discusses a case study for degradation monitoring of a structural system followed by concluding remarks in Section \ref{sec:conclusion}.

%%% -----------------------------------------------------------------
\section{Background} \label{bg}

\subsection{Dynamic Bayesian Network (DBN)}
\label{subsec:dbn}

\begin{figure}[h]
    \centering
    \includegraphics[scale=0.43]{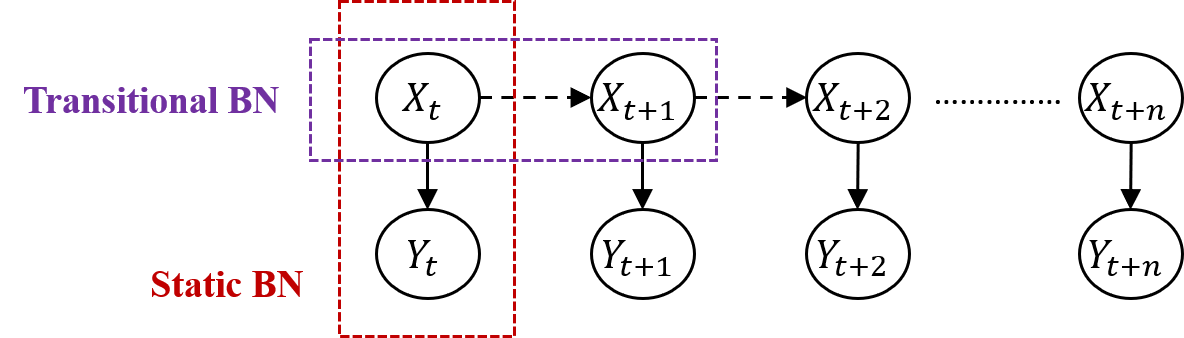}
    \caption{A schematic of a DBN for modeling time-dependent systems}
    \label{fig:dbn_ex}
\end{figure}

A DBN falls under the state-space modeling paradigm, the state of the system is modeled using a set of unobservable `state' variables, which are inferred through data available on a set of `observation’ variables. For modeling through a DBN, the continuous time is discretized into discrete time steps, and the system is analyzed over these discrete time steps. Fig. \ref{fig:dbn_ex} shows a schematic of a DBN for modeling time-dependent systems. 

In Fig. \ref{fig:dbn_ex}, $X_t$ and $Y_t$ are state and observation variable at time step $t$. Following our previous work, we consider a DBN as a composition of two BN: (1) a static BN that models the stochastic relationships between variables in the same time step, and (2) a transitional BN that models stochastic relationships across two consecutive time steps. Fig. \ref{fig:dbn_ex} shows temporal relationships between the state variables (dashed arrows) and relationships within the same time steps are represented using solid arrows. Here, the temporal relationships are present between two consecutive time steps only (also referred to as Markov property); this is commonly assumed when modeling using a DBN \cite{murphy2002dynamic}. The conditional probabilities in the static and transitional BNs ($P(Y_t|X_t)$ and $P(X_{t+1}|X_t)$) remain the same and do not change over time. 

\textit{Analysis using a DBN}: Given data on $Y_t$, we obtain the posterior probabilities of $X_t$ at time $t$ through Bayesian inference analysis using the static BN. Using the posterior probabilities of $X_t$, we obtain the prior probabilities of state variables at time $t+1$ (i.e. $X_{t+1}$) using the transitional BN.  

\subsection{Quantum Bayesian Network (QBN)}
\label{subsec:qbn}

In this section, we provide a brief description regarding gate-based circuit representation of quantum Bayesian networks (QBNs). Here, we discuss the Compositional Quantum Bayesian Network (C-QBN) approach developed as our prior work \cite{borujeni2021quantum} to represent any generic discrete QBN. We follow three steps for QBN circuit representation. 

The first step is map each variable in a BN to one or more qubits. A qubit can be in two states. If a random variable in a BN has more than two states, then multiple qubits need to be used to represent a BN variable. If $n_s$ represents the number of random variable states, then the number of qubits required can be calculated as $n_q = \lceil log_2 n_s \rceil$. 

In the second step, we map the marginal and conditional probabilities of various nodes in a BN to probability amplitudes of corresponding qubits. Finally, the desired probabilities are realized by implementing (controlled) rotation gates. In the C-QBN approach, a quantum circuit is obtained by composing different blocks of gates, each block of gates correspond to realizing marginal/conditional probabilities associated with a BN variable.

\begin{figure}[h]
\begin{center}
\hspace{5mm}\Qcircuit @C=0.4em @R=0.5em {
\lstick{} & \qw & \gate{R_Y(\theta_X)} & \ctrl{1} & \gate{X} & \ctrl{1} & \gate{X} &\qw 
\inputgroup{1}{1}{0em}{X_t}\\
\lstick{} & \qw & \qw & \gate{R_Y(\theta_{Y,\Ket{1}})} & \qw  & \gate{R_Y(\theta_{Y,\Ket{0}})}&\qw &\qw
\inputgroup{2}{2}{0em}{Y_{t}}
}
\end{center}
    \caption{A Quantum Bayesian network circuit with two nodes(qubits)}
    \label{fig:twonodeBN}
\end{figure}
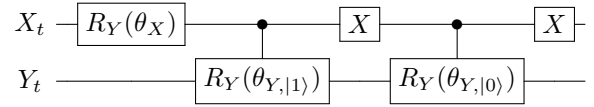

For illustration, let us consider the static BN at time $t$ in Fig. \ref{fig:dbn_ex}, which has two variables $X_t$ and $Y_t$. Let us assume that each of $X_t$ and $Y_t$ has two states. Since they have two states, we use one qubit to represent each of them. Let $X_t=0,1$ and $Y_t=0,1$ represent the two states of $X_t$ and $Y_t$. The 0 and 1 states are mapped to $\Ket{0}$ and $\Ket{1}$ states of their respective qubits. Fig. \ref{fig:twonodeBN} shows the QBN ciruit for the static BN with $X_t$ and $Y_t$.

First, we implement a rotation gate ($R_Y$) to realize the marginal probabilities of $X_t$. With the implementation of $R_Y(\theta_X)$ gate, a qubit in an initial state of $\Ket{0}$ gets transformed to $\Ket{0} \longrightarrow \cos\big(\frac{\theta_X}{2}\big)\Ket{0}$ + $\sin\big(\frac{\theta_X}{2}\big)\Ket{1}$ with $\cos^2\big(\frac{\theta_X}{2}\big)$ and $\sin^2\big(\frac{\theta_X}{2}\big)$ being the probabilities of $\Ket{0}$ and $\Ket{1}$ states respectively. Thus the value of the rotation angle can be calculated as $\theta_X = 2 \arctan\Big(\sqrt{\frac{P(X_t=1)}{P(X_t=0)}}\Big)$. 

Since $Y_t$ is a child node of $X_t$, we have a set of probabilities for $Y_t$ conditioned on the value of $X_t$. These conditional probabilities are realized through controlled rotation gates. $\theta_{Y,\Ket{1}}$ and $\theta_{Y,\Ket{0}}$ represent the rotation angles that need to implemented to realize the conditional probabilities associated with $X_t=1$ and $X_t=0$ respectively.

When a variable in a BN has more than two states we need more than one qubit to represent it in the circuit. In that case instead of applying a single-qubit rotation, a transformation $U$ is applied on the set of qubits for that variable. 

\begin{figure*}[!ht]
\begin{center}
\hspace{1mm}\Qcircuit @C=1em @R=0em {
\lstick{\Ket{0}}& \qw & \multigate{4}{U} & \qw & \qw \\
\lstick{\Ket{0}}& \qw & \ghost{U} & \qw & \qw \\
\lstick{\Ket{0}}& \qw & \ghost{U} & \qw & \qw  & \push{=} \\
& \cdots & \nghost{U} & \cdots & \\
\lstick{\Ket{0}}& \qw    & \ghost{U} & \qw & \qw} \hspace{20mm}\Qcircuit @C=1em @R=0em {
\lstick{q_0=\Ket{0}}& \gate{R_Y(\theta)} & \ctrl{1} & \gate{X} & \ctrl{1} & \gate{X} & \qw \\
\lstick{\Ket{0}}& \qw & \multigate{3}{U_{q_0 = \Ket{1}}} & \qw & \multigate{3}{U_{q_0 = \Ket{0}}} &\qw & \qw\\
\lstick{\Ket{0}}& \qw & \ghost{U_{q_0 = \Ket{1}}} & \qw & \ghost{U_{q_0 = \Ket{0}}}&\qw& \qw \\
& \cdots & \nghost{U_{q_0 = \Ket{1}}} & \cdots & \nghost{U_{q_0 = \Ket{0}}} & \cdots\\
\lstick{\Ket{0}}& \qw & \ghost{U_{q_0 = \Ket{1}}} & \qw & \ghost{U_{q_0 = \Ket{0}}}&\qw& \qw}
\end{center}
\caption{Decomposition of $U$ gate implemented on multiple qubits to represent a variable with more than two states}
\label{fig:Utrans}
\end{figure*}
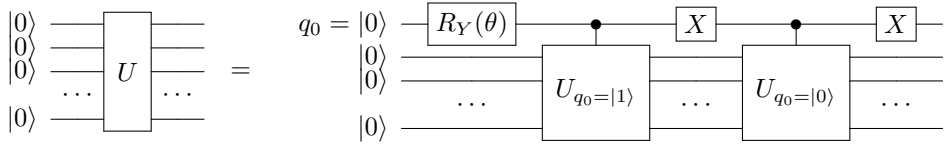

Inside that gate $U$, proper rotations will transform the qubits to obtain desirable probabilities. Fig. \ref{fig:Utrans} illustrates such a situation if variable $X_t$ had more than two states. In such a case, $U$ needs to be decomposed into a set of single qubit and CNOT basis gates. We adopt the sequential decomposition approach that was developed by Borujeni et al \cite{borujeni2021quantum} to represent a variable with more than two states.

%%% -----------------------------------------------------------------
\section{Dynamic Quantum Bayesian Network (DQBN)}
\label{sec:method}

\begin{figure}[!ht]
    \centering
    \includegraphics[scale=0.45]{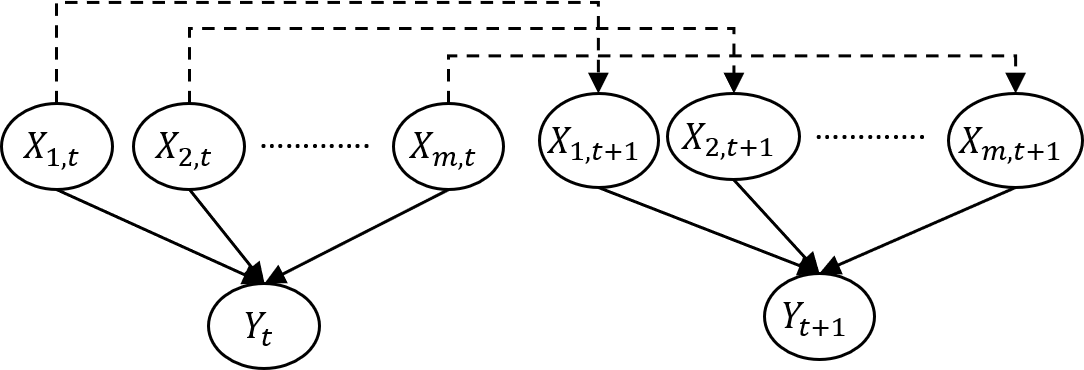}
    \caption{A dynamic Bayesian network for a node with $m$ parents}
    \label{fig:dbnnominal}
\end{figure}

Let $X_1,X_2, ..., X_m $ and $Y$ be the set of state variables and $Y$ is an observation variable. Let $X_{1,t},X_{2,t}, ..., X_{m,t}$ and $Y_t$ be the state and observation variables at time $t$. As mentioned in section \ref{bg}, we consider that the Markov assumption holds good for our DBN, meaning that the state variables in each time step, only depend on their state in the previous time step. This leads us to a two time step DBN as shown in Fig. \ref{fig:dbnnominal}.

Following the notation in Fig. \ref{fig:dbn_ex}, the dashed arrows represent the temporal relationships (between $X_{i,t}$ and $X_{i,t+1}$, $i = 1,2,\dots m$) while the solids arrows represent (between $X_{i,t}$ and $Y_t$) relationships within the same time step. We first discuss below the analysis steps performed using a DBN, and later we present the equivalent quantum analysis.

\textbf{Step 1:} Given the static BN at time $t$ (including the marginal prior probabilities of $X_{i,t}$ and conditional probability table of $Y_t$), we obtain the posterior probabilities of $X_{i,t}$ through Bayesian inference analysis using observation data on $Y_t$.

\textbf{Step 2:} Using the posterior probabilities of $X_{i,t}$, we obtain the prior probabilities of $X_{i,t+1}$ by performing forward prediction using the temporal conditional probabilities in the transitional BNs. Note that in the case, there are $m$ transitional BNs, one corresponding to each $X_{i,t}$. As these $m$ transitional BNs are independent to each other, these analyses can be performed in parallel.

We discuss below the above two steps on the quantum computing framework.

\textbf{QStep 1:} Before performing the Bayesian inference analysis, we need to obtain the circuit representation of the static BN. Following the discussion in Section \ref{subsec:qbn}, we briefly discuss the representation of the static BN. 

\textit{Static BN representation:} Let $A$ denote the quantum circuit that represents our BN and $\ket{\psi_0}$ be the state of the system after applying the unitary operator $A$ on the qubits.
\begin{equation}
\Qcircuit @C=0.5em @R=0.7em {
\lstick{\Ket{0^{\otimes {(m+1)}}}} & \qw & \gate{A} & \qw &\qw &\meter&\qw & \qw && && = \Ket{\psi_0}}
\label{eq:method1}
\end{equation}

The circuit for the static BN in Fig.\ref{fig:dbnnominal} within one time step, contains $m$, $R_Y$ rotations for $X_{1,t},X_{2,t}, \dots, X_{m,t} $ and the number of different combinations of their states identifies the number of controlled rotations to be implemented to realize the conditional probabilities associates with $Y_t$. For a special case where all the variables are binary, $A$ contains $2^m$ $C^mR_Y$ rotations (with $m$ control qubits and one target qubit). The implementation of $C^mR_Y$ when $m \ge 2$ requires $m-1$ ancilla qubits \cite{borujeni2021quantum}. 

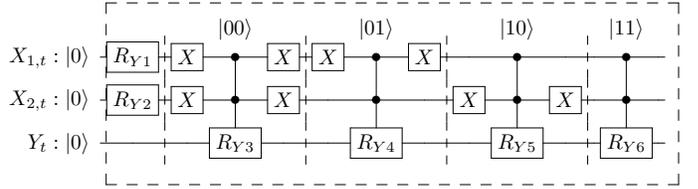
\begin{figure}[!h]
\begin{center}
\scalebox{0.8}{
\hspace*{12mm} \Qcircuit @C=0.3em @R=0.6em { 
 && & & \mbox{$\ket{00}$} & && & \mbox{$\ket{01}$}& && & \mbox{$\ket{10}$} & && & \mbox{$\ket{11}$} & & & &  \\
\lstick{X_{1,t}:\Ket{0}}  &\gate{R_{Y1}}\barrier[0em]{0}&\qw &\gate{X}&  \ctrl{1} &\gate{X} \barrier[0em]{0}&\qw&\gate{X}& \ctrl{1} &\gate{X} \barrier[0em]{0}&\qw&\qw& \ctrl{1} &\qw \barrier[0em]{0}&\qw&\qw&  \ctrl{1} &\qw& \qw  \\
\lstick{X_{2,t}: \Ket{0}} &\gate{R_{Y2}}\barrier[0em]{0}&\qw &\gate{X}&  \ctrl{1} &\gate{X} \barrier[0em]{0}&\qw&\qw&  \ctrl{1} &\qw \barrier[0em]{0}&\qw&\gate{X}&  \ctrl{1} &\gate{X} \barrier[0em]{0}&\qw&\qw&  \ctrl{1} &\qw& \qw  \\
\lstick{Y_t:\Ket{0}} &\qw \barrier[0em]{0}&\qw &\qw& \gate{R_{Y3}} &\qw \barrier[0em]{0}&\qw&\qw&  \gate{R_{Y4}} &\qw\barrier[0em]{0}&\qw&\qw& \gate{R_{Y5}} &\qw \barrier[0em]{0}&\qw&\qw&  \gate{R_{Y6}} &\qw&\qw \gategroup{1}{2}{4}{17}{2.5em}{--} }}
\end{center}
\caption{Illustrative QBN with three binary variables (two state variables and one observation variable)}
\label{fig:example}
\end{figure}

Fig. \ref{fig:example} shows the circuit for a simple case of $m=2$ for binary variables $X_{1,t},X_{2,t} \text{ and } Y_t$. In this circuit, $R_{Y1} \text{and } R_{Y2} $ are applied for marginals and the rest of the circuit is composed of 4 section contains $CCR_Y$ rotations for the four possible combinations of $X_{1,t} \text{ and } X_{2,t} $ i.e. $\ket{00}, \Ket{01}, \Ket{10} \text{and } \Ket{11}$. This whole circuit as shown inside the dashed box represents unitary operator $A$. \\

\textit{Bayesian inference:} The estimation of posterior probabilities through Bayesian inference is carried out using the quantum amplitude amplification algorithms through Grover's operator. The posterior probabilities can be calculated as 

\begin{equation}
\label{eqn:post}
\begin{aligned}
     & P(X_{1,t}, X_{2,t}, \dots X_{m,t}|Y_t=y_t)\\
     &  = \dfrac{P(X_{1,t}, X_{2,t}, \dots X_{m,t}, Y_t=y_t)}{\sum_{X_{1,t}, X_{2,t}, \dots X_{m,t}} P(X_{1,t}, X_{2,t}, \dots X_{m,t}, Y_t=y_t)} \\
     &  = \dfrac{n(X_{1,t}, X_{2,t}, \dots X_{m,t}, Y_t=y_t)}{\sum_{X_{1,t}, X_{2,t}, \dots X_{m,t}} n(X_{1,t}, X_{2,t}, \dots X_{m,t}, Y_t=y_t)}
\end{aligned}
\end{equation}

In Eq. \ref{eqn:post}, $y_t$ is the observation data point available on $Y_t$. $P(.)$ is the probability function while $n(.)$ represents the function that provides the counts of a state. Here, the Grover's operator for amplitude amplification is shown in Eq. \ref{eq:GroverOp} \cite{low2014quantum}.

\begin{equation}
    G =    {S_e} A^{\dagger} S_0 A
\label{eq:GroverOp}
\end{equation} 

In Eq. \ref{eq:GroverOp}, ${G}$ denotes the Grover's operator, ${A}$ is the QBN circuit (corresponding to the static BN), $S_0$ is the zero phase shift (reflection operator), ${A}^{\dagger}$ is the conjugate transpose of ${A}$, and ${S_e}$ is the phase oracle defined below in Eq. \ref{eqn:se}.

\begin{equation}
\label{eqn:se}
\begin{aligned}
  &  S_e : \ket{x} \xrightarrow{} (-1)^{f(x)}\ket{x} , \text{where}:\\
  & f(x) = \begin{cases}
    1 , \text{if } x \text{ is a good state }\\
    0 , \text{otherwise}\\
    \end{cases}
    \\
    \end{aligned}
\end{equation}\\

In Eq. \ref{eqn:se}, $\Ket{x}$ correspond to a quantum state in the QBN circuit. Combining Eq. \ref{eq:method1} with the Grover operator (Eq. \ref{eq:GroverOp}), we have

\begin{equation}
\Qcircuit @C=0.5em @R=0.7em {
\lstick{\Ket{0^{\otimes {(m+1)}}}} & \qw & \gate{A} & \qw &\gate{G}&\qw & \qw }
\label{eq:method2-1}
\end{equation}
or
\begin{equation}
\Qcircuit @C=0.5em @R=0.7em {
\lstick{\Ket{\psi_0}} & \qw & \gate{S_e} & \qw &\gate{A^{\dagger}} & \qw &\gate{S_0} & \qw& \gate{A} & \qw& \meter & \qw &&& & = \Ket{\psi_1} }
\label{eq:method2-2}
\end{equation}

The Grover operator here, amplifies the probabilities of the good states, which in our case, are the states that are inferred. Depending on the number of good states, we implement multiple Grover iterations as shown in Eq. \ref{eq:method3}. 

\begin{equation}
\Qcircuit @C=0.5em @R=0.9em { &&&&& \mbox{$k$ iterations} &&&&&&&\\
\lstick{\Ket{\psi_0}} & \qw & \gate{S_e} & \qw &\gate{A^{\dagger}} & \qw &\gate{S_0} & \qw& \gate{A} & \qw& \meter & \qw &&& & = \Ket{\psi_{k}} \gategroup{2}{3}{2}{9}{0.7em}{--}}
\label{eq:method3}
\end{equation}

%%%% 

\textbf{QStep 2:} After obtaining the posterior probabilities of the state variables, we calculate their prior probabilities in the next time step using the transitional BN. In Fig \ref{fig:dbnnominal}, we have $m$ transitional BNs, one corresponding to each $X_{i,t}$. Fig. 6 shows the quantum circuit of a transitional BN.

\begin{figure}[!h]
\begin{center}
\hspace{2.5em}\Qcircuit @C=0.1em @R=0.05em @!R 
{\lstick{X_{i,t}} & \qw & \gate{R_Y(\theta_{i,t})} & \qw & \ctrl{1} &\qw & \gate{X} & \qw & \ctrl{1}& \qw & \gate{X} & \qw\\
\lstick{X_{i,t+1}} & \qw &\qw  & \qw & \gate{R_Y(\theta_{i,t, \Ket{1}})} & \qw &\qw & \qw & \gate{R_Y(\theta_{i,t, \Ket{0}})} & \qw & \qw& \qw}  
\end{center}
\label{fig:transqbn}
\caption{Illustrative transitional QBN of a state variable over two time steps}
\end{figure}
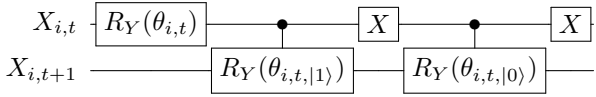

In Fig. 6, $\theta_{i,t}$ is the rotation angle that corresponds to the posterior probabilities of $X_{i,t}$, and $\theta_{i,t, \Ket{1}}$ and $\theta_{i,t, \Ket{0}}$ correspond to the rotation angles to be implemented to realize the probabilities of $X_{i,t+1}$ when conditioned on $X_{i,t}=1$ and $X_{i,t}=0$. It should be noted that the prior probabilities of $X_{i,t+1}$ can be obtained by simulating the transitional QBN without any need for Grover operators and inference analysis.

Since the conditional probabilities in the static and transitional BNs remain the same across multiple time steps, the architectures of the static and transitional QBNs (in Figs. 5,6) also remain the same. In both the QBN circuits, the rotation angles associated with the marginal probabilities vary over time (e.g. $\theta_{i,t}$ in Fig. 6; $R_{Y1}$ and $R_{Y2}$ in Fig. 5).

\section{Case Study, Results, and Discussion}
\label{sec:case}

\begin{figure}[ht]
    \centering
    \includegraphics[scale=0.42]{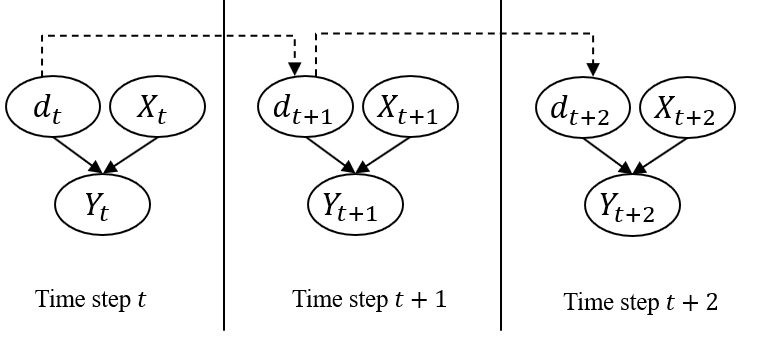}
    \caption{The dynamic Bayesian network with 3nodes over three time steps}
    \label{fig:dbnmain}
\end{figure}

In this section, we demonstrate the proposed DQBN approach for health monitoring of a simulated structural system, where $X_t$ and $Y_t$ represent the load and response of a structural system respectively. In addition to the load, the response also depends on any structural degradation (e.g. cracks); this is denoted as $d_t$. The motivation for this example was derived from work done by Bartram \cite{bartram2013system} and Li et al \cite{li2017dynamic}, where DBNs were used for health management and prognostics of mechanical and aerospace systems respectively. The DBN of this example is given in Fig. \ref{fig:dbnmain}.

Both the load and response variables ($X_t$, $Y_t$) are discretized into three levels - \{Low, Medium, High\}and denoted as \{0, 1, 2\}. The deterioration variable $d_t$ is discretized into two levels - \{Minor, Major\} and denoted as \{0, 1\} respectively. The marginal probabilities of $d_t$ at time $t=0$ are given as $P(d_0=0)=0.95$ and $P(d_0=1)=0.05$. Similarly, the probabilities of $X_t$ at time $t=0$ are given as $P(X_0=0)=0.2$, $P(X_0=1)=0.5$ and $P(X_0=2)=0.3$. The probabilities of $Y_t$ conditioned on $d_t$ and $X_t$ are given in Table \ref{tab:cpt}.

\begin{table}[h]
    \centering
    \caption{Conditional probabilities of response dependent of degradation and load}
    \begin{tabular}{|c|c|c|c|c|c|c|}
         \hline
         & (0,0) & (0,1) & (0,2) & (1,0) & (1,1) & (1,2)\\
         \hline
         $P(Y_t=0|d_t, X_t)$& 0.8 & 0.75 & 0.65 & 0.15 & 0.05 & 0 \\
         \hline
         $P(Y_t=1|d_t, X_t)$& 0.15 & 0.18 & 0.23 & 0.55 & 0.6 & 0.35 \\
         \hline
         $P(Y_t=2|d_t, X_t)$& 0.05 & 0.07 & 0.12 & 0.3 & 0.35 & 0.65 \\
         \hline
    \end{tabular}
    \label{tab:cpt}
\end{table}

Engineering systems age and their performance naturally degrades over time. Here, the degradation is modeled through a transitional conditional probability distribution across two steps following the Markov property, i.e., the degradation in the current time step is dependent on the degradation in the previous time step. The conditional distribution of $d_{t+1}$ dependent on $d_t$ is given in Table \ref{tab:tcpt}.

\begin{table}[h]
    \centering
    \caption{Conditional probability relationship of degradation across two time steps}
    \begin{tabular}{|c|c|c|}
        \hline
         & $d_t$ = 0 & $d_t$ = 1 \\
         \hline
         $P(d_{t+1}= 0)$ & 0.9 & 0 \\
         \hline
         $P(d_{t+1}= 1)$ & 0.1 & 1 \\
         \hline
    \end{tabular}
    \label{tab:tcpt}
\end{table}

We perform two types of analysis through this case study.
\begin{enumerate}
    \item Track probability distribution of degradation over time
    \item Compare the solution accuracy between classical analysis, IBM Qiskit simulator and IBM Q hardware. 
\end{enumerate}

\textbf{Modeling DQBN}: As discussed in Section \ref{sec:method}, we decompose the DBN into two BNs: (1) Static BN, and (2) Transitional BN. We model the joint distribution between $d_t$, $X_t$ and $Y_t$ through the static BN at each time step, and we  model the joint distribution between the degradation variables across two consecutive time steps.  Here, we first discuss the circuit representations of static and transitional BNs at time $t=0$.

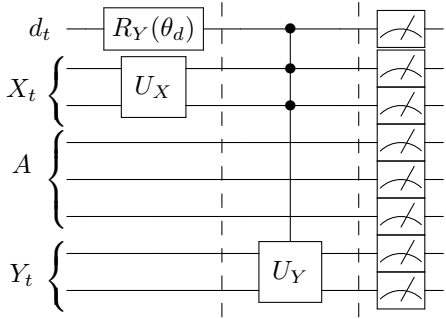
\begin{figure}[h]
\begin{center}
\hspace{1mm}\Qcircuit @C=0.7em @R=0em {
\lstick{} & \qw & \gate{R_Y(\theta_d)} \barrier[0em]{7}& \qw & \qw & \ctrl{1} & \qw \barrier[0em]{7}& \qw & \meter & \qw 
\inputgroup{1}{1}{0em}{d_t} \\
\lstick{}& \qw & \multigate{1}{U_{X}} & \qw & \qw & \ctrl{1} & \qw & \qw &\meter & \qw \\
\lstick{}& \qw & \ghost{U_{X}} & \qw & \qw& \ctrl{4} & \qw & \qw &\meter & \qw 
\inputgroupv{2}{3}{0.8em}{0.8em}{X_t} \\
\lstick{}& \qw & \qw & \qw & \qw & \qw & \qw & \qw &\meter & \qw \\
\lstick{}& \qw & \qw & \qw & \qw & \qw & \qw & \qw &\meter & \qw \\
\lstick{}& \qw & \qw & \qw & \qw & \qw& \qw & \qw & \meter & \qw 
\inputgroupv{4}{6}{0.8em}{0.8em}{A} \\
\lstick{}& \qw & \qw & \qw & \qw & \multigate{1}{U_{Y}} & \qw & \qw &\meter & \qw \\
\lstick{}& \qw & \qw & \qw & \qw & \ghost{U_{Y}} & \qw & \qw &\meter & \qw 
\inputgroupv{7}{8}{0.8em}{0.8em}{Y_t}
}
\end{center}
\caption{Schematic of the static BN}
\label{fig:staticBN}
\end{figure}

\underline{\textit{Static BN at time $t=0$:}} A schematic representation of the static BN is shown in Fig. \ref{fig:staticBN}. Since $X_t$ and $Y_t$ each has three levels, we use two qubits to represent each of these two variables. We use one qubit to represent the two levels of $d_t$, totaling to five qubits.

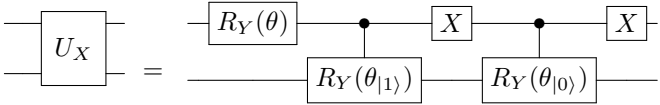
\begin{figure}[!ht]
\begin{center}
\hspace{1mm}\Qcircuit @C=0.7em @R=1em {
\lstick{}& \qw & \multigate{1}{U_X} & \qw & \\
\lstick{}& \qw & \ghost{U_X} & \qw & \push{=}} \hspace{2mm} 
\Qcircuit @C=0.4em @R=0.5em {
   \lstick{}& \qw&\gate{R_Y(\theta)} & \ctrl{1}& \gate{X} & \ctrl{1} & \gate{X} & \qw\\
   \lstick{}& \qw& \qw & \gate{R_{Y}(\theta_{\Ket{1}})} & \qw & \gate{R_{Y}(\theta_{\Ket{0}})} & \qw & \qw}
\end{center}
\caption{Circuit representation of a two-qubit gate using single qubit and controlled rotations}
\label{fig:twoqubit}
\end{figure}

Since $P(d_0=0)=0.95$ and $P(d_0=1)=0.05$, the rotation angle to be implemented to realize these probabilities can be calculated as $\theta_d = 2\arctan\Big(\sqrt{\frac{P(d_0=1)}{P(d_0=0)}}\Big) = 0.451$. The variable $X_0$ (i.e. $X_t$ at $t=0$) has three states 0,1, and 2. We represent these three states using two qubits. We map the three states to $\ket{00}$, $\ket{01}$, and $\ket{10}$ states respectively. As discussed in Section \ref{bg}, we decompose the two-qubit $U_X$ gate into a set of single qubit and CNOT gates shown in Fig. \ref{fig:twoqubit}.

In Fig. \ref{fig:twoqubit}, $R_Y(\theta)$ is the single-qubit rotation gate implemented on the first qubit, $\theta_{\Ket{1}}$ and $\theta_{\Ket{0}}$ are the conditional rotation angles implemented on the second qubit with the first qubit as the control qubit. Here, $\theta = 2\arctan\Big(\sqrt{\frac{P(\Ket{10}) + P(\Ket{11})}{P(\Ket{00}) + P(\Ket{01})}}\Big) = 1.159$. Similarly, $\theta_{\Ket{1}}$ and $\theta_{\Ket{0}}$ can be calculated as $\theta_{\Ket{1}} = 2\arctan\Big(\sqrt{\frac{P(\Ket{11})}{P(\Ket{10})}}\Big) = 0$ and $\theta_{\Ket{0}} = 2\arctan\Big(\sqrt{\frac{P(\Ket{01})}{P(\Ket{00})}}\Big) = 2.014$. Depending on the values of $d_t$ and $X_t$, we have several conditional probabilities as given in Table \ref{tab:cpt}, which are represented using conditional rotations over two qubits (shown in Fig. \ref{fig:staticBN}). As there are six combinations of $d_t$ and $X_t$, we will have six different $U_Y$ gates. 

Similar to $U_X$, $U_Y$ can also be decomposed into single qubit and controlled rotation gates. As shown in Fig. \ref{fig:staticBN}, we need to implement three-qubit (one $d_t$ and two of $X_t$) controlled rotation over two qubits in order to realize the conditional probabilities of $Y_t$. Decomposition of $U_Y$ itself requires a controlled rotation; therefore, implementation of $C^3U_Y$ gate requires four-qubit (three for $d_t$ and $X_t$, and one for decomposition of $U_Y$ as shown in Fig. \ref{fig:twoqubit}) controlled rotation. We use three ancilla qubits to realize the four-qubit controlled rotation. In total, the circuit requires eight qubits; five qubits to represent the three variables and three ancilla qubits to implement the controlled rotation gates. 

It should be noted that the rotation angle $\theta_d$ changes at each time step and the rest of the static BN remains the same. Note that the marginal probabilities of $X_t$ are do not change with time. Therefore, we can use the same static BN (with the change in $\theta_d$) in each time step without the need to construct a different static BN circuit at each time step.

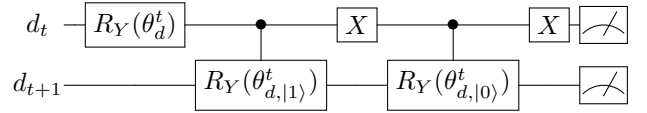
\begin{figure}[h]
\begin{center}
\hspace{5mm}\Qcircuit @C=0.4em @R=0.5em {
\lstick{} & \qw & \gate{R_Y(\theta_d^t)} & \ctrl{1} & \gate{X} & \ctrl{1} & \gate{X} & \meter 
\inputgroup{1}{1}{0em}{d_t}\\
\lstick{} & \qw & \qw & \gate{R_Y(\theta_{d,\Ket{1}}^t)} & \qw  & \gate{R_Y(\theta_{d,\Ket{0}}^t)}&\qw & \meter 
\inputgroup{2}{2}{0em}{d_{t+1}}
}
\end{center}
    \caption{Schematic of the transitional BN across two consecutive steps}
    \label{fig:transBN}
\end{figure}

\underline{\textit{Transitional BN across two time steps:}} A schematic representation of the transitional BN with degradation variables across two consecutive time steps is given in Fig. \ref{fig:transBN}.

In Fig. \ref{fig:transBN}, $\theta_d^t$ represents the angle to realize the \underline{posterior} probabilities of $d_t$ (and not the prior probabilities). The posterior probabilities can be obtained using data on $Y_t$ and the quantum amplitude amplification algorithm. Using the posterior probabilities of $d_t$ at time step $t$, we calculate the prior probabilities of $d_t$ at time $t+1$. In Fig. \ref{fig:transBN}, $\theta_{d,\Ket{1}}^t$ and $\theta_{d,\Ket{0}}^t$ are rotation angles associated with the conditional probabilities of $d_{t+1}$ when $d_t=1$ and $d_t=0$ respectively. Similar to the static BN circuit, the transitional BN circuit does not change except for the change in $\theta_d^t$ as the transitional conditional probabilities remain the same across any two time steps.

\textbf{Degradation estimation:} So far, we discussed about the circuit representation of static and transitional BNs. Here, we will discuss the estimation of state variable (degradation; $d_t$) using data on observation variables (response variables $Y_t$ and $X_t$). We will demonstrate the proposed framework for five time steps. The observation data of $Y_t$ across these five steps is assumed as \{0, 1, 1, 1, 2\}. We also consider the observation data for $X_t$ at these five time steps as \{1,0,2,1,2\}.  In the first time step, the observation of $Y_t$ is 0 and the observation of $X_t$ is 1. As discussed, $Y_t=0$ is represented using two qubits as $\Ket{00}$ and $X_t=1$ would be $\Ket{01}$. For posterior distribution of $d_t$, we need to calculate $P(d_t=0|Y_t=0, X_t=1)$ and $P(d_t=1|Y_t=0, X_t=1)$.

The estimation of degradation variable through quantum amplification requires us to build an Oracle. In the Oracle, we fix the two qubits related to $Y_t$ at $\Ket{0}$ (resulting in $\Ket{00}$ state), fix the two qubits related to $X_t$ at $\Ket{0}$ and $\Ket{1}$ respectively (resulting in $\Ket{01}$ state) and
the three ancilla qubits at $\Ket{0}$ state. Ideally, the ancilla qubits should be in $\Ket{0}$; however, the presence of experimental noise cause the ancilla qubits to not be in the exact $\Ket{0}$ states. Fixing the ancilla qubits at $\Ket{0}$ results in a more effective estimation of the state (degradation) variable. Using this oracle, we implement the Grover rotations to calculate the posterior probabilities of $d_t$.

Of the eight qubits, we fix the values of seven qubits (two relating to $Y_t$, two relating to $X_t$ and three ancilla qubits) while the remaining qubit can either be in $\Ket{0}$ or $\Ket{1}$ state. We ran the analysis with 8192 shots and obtain the number of counts when the qubit associated with $d_t$ is in $\Ket{0}$ and $\Ket{1}$ states. If $n_0$ and $n_1$ represent the amount of such counts, then the posterior probabilities can be calculated as $P(d_t=0|Y_t=0, X_t=1) = \frac{n_0}{n_0 + n_1}$ and $P(d_t=1|Y_t=0, X_t=1) = \frac{n_1}{n_0 + n_1}$. Three iterations of Grover's operator can give us the optimal solution while adding more iterations worsens the results. These posterior probabilities are then used to calculate the prior probabilities in the next time step using the transitional BN circuit in Fig. \ref{fig:transBN}.

\begin{figure}
    \centering
    \includegraphics[scale=0.68]{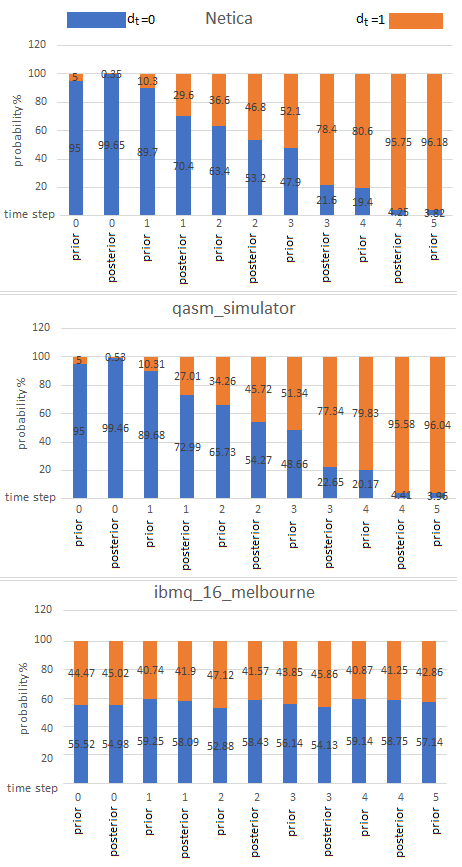}
    \caption{Prior and posterior distribution of $d_t$ over five consecutive time steps using Netica, qasm\_simulator, and ibmq\_16\_melbourne }
    \label{fig:barplot}
\end{figure}

\textbf{Discussion:}
The results of the DQBN analysis are provided in Fig.\ref{fig:barplot}. The three bar plots provide the prior and posterior probabilities of degradation variable $d_t$ across five time steps. It could be observed that the qasm\_simulator results closely match the classical results (through Netica software \cite{Netica}). The root mean squared (RMS) error in the prior and posterior probabilities of $d_t=0$ between Qiskit simulator and classical implementation is calculated as 1.192\%. The RMS error for IBM Melbourne is equal to 34.66\%. We believe the discrepancy in the results between IBM hardware and Qiskit simulator is due to the inherent hardware (gate) errors, decoherence, and the large depth of the circuits. For example, the depth of the \underline{\textit{transpiled}} static QBN circuit at time $t=0$ is 946. Even though the static BN has three variables $d_t$, $X_t$ and $Y_t$, the QBN representation required eight qubits (including three ancilla qubits) and contained 769 CNOT gates. This is the reason for the larger depth of the transpiled circuit. 

%%% ------------------------------------------------
\section{Conclusion}
\label{sec:conclusion}

This paper described state-space modeling of time-dependent systems using Dynamic Quantum Bayesian networks (DQBN). DQBN are extensions of Quantum Bayesian networks (QBNs) for modeling dynamic systems. We considered a DQBN as a composition of two QBNs: a static QBN that describes the relationships between variables at any given time, and a transitional QBN that describes the relationships between variables across two consecutive time steps. At any given time step, the posterior probabilities of unobserved state variables are estimated using data on observation variables using quantum amplitude amplification algorithms through Grover iterations. The posterior probabilities of the state variables are then used to obtain their prior probabilities in the next time step through transitional QBN simulation. In this way, through repeated evaluation of static and transitional QBNs, we can track the performance of time-dependent systems.  This paper demonstrated the proposed framework for degradation monitoring of a structural system. We performed the analysis on the IBM Qiskit simulator and IBM hardware and compared their performance against classical implementation (through Netica software). We observed that the IBM simulator results were close to the classical results whereas the results from IBM hardware (Melbourne device) were error-prone, noisy, and therefore, unreliable. 

As future work, we will investigate the scalability of the proposed framework for high-dimensional systems (increasing the number of variables and increasing the number of states of each variable). Moreover, we will also incorporate gate and measurement error mitigation strategies to improve the solution performance on the IBM hardware. We will also investigate other quantum inference algorithms for state estimation such as quantum Metropolis algorithm \cite{temme2011quantum} and variational inference methods \cite{benedetti2021variational}. 

\section*{Acknowledgement}

We acknowledge the use of IBM Quantum services for this work. The views expressed are those of the authors, and do not reflect the official policy or position of IBM or the IBM Quantum team.

%%% ------------------------------------------------
\bibliographystyle{IEEEtran}
\bibliography{references}

\end{document}